\documentclass{article}
\usepackage{spconf,graphicx,cite}

\usepackage{array,amssymb,amsmath,amsfonts}
\usepackage{subfigure}
\usepackage{multirow}
\usepackage{math}
\usepackage{tikz}
\usetikzlibrary{arrows,automata,shapes}
\usepackage{bm,color}% bold math

\newtheorem{theorem}{Theorem}

\title{Towards a Generalization of Relative Transfer Functions to More Than One Source}
\name{Antoine Deleforge${}^*$, Sharon Gannot${}^\dagger$, Walter Kellermann${}^*$\thanks{The research leading to these results has received funding from the
European Union’s Seventh Framework Programme (FP7/2007-2
013) under grant agreement n${}^\circ$ 609465 (project EARS).}\vspace{-2mm}}
\address{${}^*$University of Erlangen-Nuremberg, Germany\\
         ${}^\dagger$Bar-Ilan University, Israel}

\begin{document}

\maketitle
\begin{abstract}
We propose a natural way to generalize relative transfer functions (RTFs) to more than one source. We first prove that such a generalization is not possible using a single multichannel spectro-temporal observation, regardless of the number of microphones. We then introduce a new transform for multichannel multi-frame spectrograms, \textit{i.e.}, containing several channels and time frames in each time-frequency bin. This transform allows a natural generalization which satisfies the three key properties of RTFs, namely, they can be directly estimated from observed signals, they capture spatial properties of the sources and they do not depend on emitted signals. Through simulated experiments, we show how this new method can localize multiple simultaneously active sound sources using short spectro-temporal windows, without relying on source separation.
\end{abstract}
\vspace{-3mm}
\section{Introduction}
\vspace{-2mm}
When sound propagates from an emitter to a receiver in a natural environment, objects along its path (\textit{e.g.}, a human or robot head, walls...) lead to reflections and reverberation. This is commonly modeled as a linear filtering and described by the convolution of the emitted signal with a so called \textit{room impulse response} (RIR). For a given room, the latter only depends on the source's spatial properties (position, orientation, directivity, diffuseness, etc.) and not on the emitted signal. The frequency domain counterparts of RIRs are \textit{acoustic transfer functions} (ATFs). Knowledge of the ATFs involved in an acoustic setup is useful in many audio signal processing applications, \textit{e.g.}, blind source separation \cite{parra2000convolutive}, beamforming \cite{affes1997signal}, sound source localization \cite{dvorkind2005time,laufer2013relative,deleforge2015acoustic}, acoustic echo cancellation \cite{benesty2001advances}.

Most existing methods to estimate ATFs rely on the synchronized emitted and received signals. However, the emitted signals are often not available, rendering the estimation of ATFs impossible without additional restrictive assumptions. For this reason, \textit{relative transfer functions} are often considered \cite{gannot2001signal}. These also capture source spatial properties and do not depend on the emitted signal, with the advantage that they can be reliably and robustly estimated directly from an observed multichannel signal \cite{markovich2009multichannel,reindl2013geometrically}. They are defined as a \textit{normalized} version of ATFs, \textit{i.e.}, the ATF at a given microphone is divided by a linear combination of the ATFs to other microphones, \textit{e.g.}, the ATF of a reference microphone. In the case of $M=2$ microphones, the log-magnitude and phase of RTFs are referred to as \textit{interaural level} and \textit{phase differences}, respectively, in the binaural hearing literature \cite{duda1997elevation,blauert2013technology}. Recently, supervised sound source localization methods making use of a training set of interaural cues \cite{deleforge2015acoustic} or of RTFs \cite{laufer2013relative} have been proposed.

In this paper, we theoretically investigate the possibility of generalizing RTFs to more than one source. Such generalizations should preserve the three key properties of RTFs, namely, they can be directly estimated from observed signals, they capture spatial properties of the sources and they do not depend on the emitted signals. We first state and prove a theorem showing that such a generalization is not possible if a single multichannel spectro-temporal observation is used. We then consider the case of multiple time observations, and propose a new transformation for multichannel, multi-frame spectrograms, \textit{i.e.}, containing several multichannel time frames in each time-frequency bin. This transformation builds on the Pl\"ucker embedding method for Grassmannian manifolds. We show that it yields a natural generalization of RTFs to multiple sources, when there are less sources than microphones. Through simulated experiments, we show how this method could be applied to the localization of multiple simultaneously active sound sources using short spectro-temporal windows, without having to separate them.

\section{Generalizing RTFs}
\vspace{-2mm}
\subsection{Single-source case and RTF properties}
\label{subsec:1source}
Let us consider a sound source emitting the spectrogram $\{s_{ft}\}_{f=1,t=1}^{F,T}\in\mathbb{C}^{F\times T}$ recorded by an $M$-microphone array, where $F$ and $T$ are the number of frequency bands and the number of time frames, respectively. Under noise-free, finite convolutive filtering assumptions and for long enough time frames, the multichannel observation $\xvect_{ft}=[x_{ft,1},\dots,x_{ft,M}]\tp\in\mathbb{C}^M$ received by $M$ microphones at frequency-time $(f,t)$ is given by
\begin{equation}
 \label{eq:1source_model}
 \xvect_{ft} = \avect_fs_{ft}
\end{equation}
where $\avect_f=[a_{f,1},\dots,a_{f,M}]\in\mathbb{C}^M$ comprises the acoustic transfer functions from the source to the $M$ microphones at frequency $f$. For a given microphone setup in a given room, $\avect_f$ solely depends on the source's \textit{spatial properties}. Therefore, (\ref{eq:1source_model}) nicely decomposes the recorded signal into a component $\avect_f$ that only captures spatial properties and a component $s_{ft}$ that only captures the source content at $(f,t)$.

If the emitted signal $s_{ft}$ is unknown, unambiguously recovering $\avect_f$ from observation $\xvect_{ft}$ is impossible, without further assumptions. However, the specific structure of Eq.~\ref{eq:1source_model} offers an attractive way to circumvent this.  Let $\nu$ be a \textit{normalizing function}, which divides an input vector by a linear combination of its entries, \textit{e.g.}, the first entry. It is then easy to check that $\nu(\xvect_{ft}) = \nu(\avect_f)$ for all $\xvect_{ft}\in\mathcal{I}$, where $\mathcal{I}\subseteq\mathbb{C}^M$ is the nonzero locus of the linear combination.  In other words, the signal term cancels out and $\rvect_{ft} = \nu(\xvect_{ft})$, when defined, captures only the spatial properties of the source. In the signal processing literature, $\rvect_f$ is referred to as a \textit{relative transfer function} (RTF) \cite{gannot2001signal}. In summary, relative transfer functions possess three key desirable properties:
\begin{itemize}
 \item[] \textit{(I)} They can be directly estimated from observed signals
 \item[] \textit{(II)} They capture spatial properties of the sound source
 \item[] \textit{(III)} They do not depend on the emitted signal
\end{itemize}
Mathematically, these three properties are verified if and only if there exists a \textit{non-constant} function $g\hspace{-0.8mm}:\mathcal{I}\rightarrow\Omega$ and a function $h$ such that (\ref{eq:1source_model})$\implies g(\xvect_{ft}) = h(\avect_f)$ for all $\xvect_{ft}\in\mathcal{I}$, where $\Omega$ is an arbitrary set and $\mathcal{I}\subseteq\mathbb{C}^M/\{\zerovect\}$.

\subsection{Instantaneous multiple-source case}
\label{subsec:nonexist}
In the case of $K$ sound sources emitting spectrograms $\{s_{ft,k}\}_{f=1,t=1}^{F,T}$ for $k=1\dots K$, model (\ref{eq:1source_model}) becomes:
\begin{equation}
 \label{eq:Ksource_model}
  \xvect_{ft} = \sum_{k=1}^K\avect_{f,k}s_{ft,k} = \Amat_{f,K}\svect_{ft}
\end{equation}
where $\svect_{ft}=[s_{ft,1},\dots,s_{ft,K}]\tp\in\mathbb{C}^K$ is the vector of emitted signals and $\Amat_{f,K}=[\avect_{f,1},\dots,\avect_{f,K}]\in\mathbb{C}^{M\times K}$ comprises the $K$ acoustic transfer functions capturing the sources' spatial properties. An interesting question is: \textit{can we generalize relative transfer functions to more than one source, while preserving properties \textit{(I)}, (\textit{II}) and (\textit{III})}? In other words, is there a non-constant function $g\hspace{-0.8mm}:\mathcal{I}\rightarrow\Omega$ and a function $h$ such that $g(\xvect_{ft}) = h(\Amat_{f,K})$ for all $\xvect_{ft}\in\mathcal{I}$? In this section, we prove that the answer is ``no'' through the following theorem:
\begin{theorem}
\label{th:nonexistence}
Let $\mathcal{I}$ be a subset of $\mathbb{C}^M/\{\zerovect\}$, $\Omega$ an arbitrary set, $g\hspace{-0.8mm}:\mathcal{I}\rightarrow\Omega$ and $h\hspace{-0.8mm}:\mathbb{C}^{M\times K}\rightarrow\Omega$ two functions and $K>1$. If for all $\Amat\in\mathbb{C}^{M\times K}$ and for all $\svect\in\mathbb{C}^K$ with $\Amat\svect\in\mathcal{I}$ we have $g(\Amat\svect) = h(\Amat)$, then $g$ is constant.
\end{theorem}
In other words, the only possible multiple-source instantaneous generalizations of RTFs are constant, which violates property \textit{(II)}.
\paragraph*{Proof of Theorem \ref{th:nonexistence}:\\ }
Let $g\hspace{-0.8mm}:\mathcal{I}\rightarrow\Omega$ and $h\hspace{-0.8mm}:\mathbb{C}^{M\times K}\rightarrow\Omega$ be two functions such that for all $\Amat\in\mathbb{C}^{M\times K}$ and for all $\svect\in\mathbb{C}^K$ with $\Amat\svect\in\mathcal{I}$ we have $g(\Amat\svect) = h(\Amat)$.\\
$\bullet$ Case $K\ge M$: Let $\Amat\in\mathbb{C}^{M\times K}$ be a fixed matrix with $M$ linearly independent columns. Then, for all $\xvect\in\mathcal{I}$, we have $\xvect=\Amat\svect$ with $\svect=\Amat\tp(\Amat\Amat\tp)^{-1}\xvect$. By definition of $g$ and $h$, we thus have $g(\xvect)=g(\Amat\svect)=h(\Amat)$ for all $\xvect\in\mathcal{I}$.  $h(\Amat)$ does not depend on $\xvect$. Therefore, $g$ is constant.\\
$\bullet$ Case $K<M$: Let $\Amat\in\mathbb{C}^{M\times K}$ be a fixed matrix with $K$ linearly independent columns. Let $E_{\Amat}$ be the column space of $\Amat$, \textit{i.e.}, the $K$-dimensional vector subspace of $\mathbb{C}^M$ defined by $E_{\Amat}=\{\Amat\svect;\svect\in\mathbb{C}^K\}$. We now prove that $g(\xvect)=h(\Amat)$ for all $\xvect\in\mathcal{I}$:
\begin{itemize}
\item[-] If $\xvect\in E_{\Amat}$, then by definition of $E_{\Amat}$ there is $\svect$ such that $\xvect=\Amat\svect$, and thus $g(\xvect)=g(\Amat\svect)=h(\Amat)$.
\item[-] If $\xvect\notin E_{\Amat}$, let $\xvect'\in E_{\Amat}\cap \mathcal{I}$. Then $\xvect$ and $\xvect'$ are linearly independent. Let $\Amat'=[\xvect,\xvect',\avect'_3,\dots,\avect'_K]\in\mathbb{C}^{M\times K}$ have $K$ linearly independent columns (note that this is only possible because $K>1$). Let $\svect=[1,0,\dots,0]\tp$ and $\svect'=[0,1,0,\dots,0]\tp$, so that $\xvect=\Amat'\svect$ and $\xvect'=\Amat'\svect'$. By definition of $g$ and $h$, we have $g(\xvect)=g(\Amat'\svect)=h(\Amat')=g(\Amat'\svect')=g(\xvect')$. Since $\xvect'\in E_{\Amat}$, we have $g(\xvect')=h(\Amat)$ and thus $g(\xvect)=h(\Amat)$.
\end{itemize}
Thus, $g(\xvect)=h(\Amat)$ for all $\xvect\in\mathcal{I}$, and $h(\Amat)$ does not depend on $\xvect$. Therefore, $g$ is constant.$\;\blacksquare$

\subsection{Multiple-frame, multiple-source case}
\label{subsec:multicase}
In this section we overcome the non-existence of an instantaneous generalization of RTFs by proposing a \textit{multi-frame generalization}. More precisely, we consider the case where $K$ rather than one observations are available along the time axis. Using the following notations:
\begin{align}
\Xmat_{ft,K} &= [\xvect_{ft},\dots,\xvect_{ft+K-1}]\in \mathbb{C}^{M\times K}, \\
\Smat_{ft,K} &= [\svect_{ft},\dots,\svect_{ft+K-1}]\in \mathbb{C}^{K\times K},
\end{align}
we obtain a multiframe version of (\ref{eq:Ksource_model}) for the time segment $[t\dots t+K-1]$:
\begin{equation}
 \label{eq:KTsource_model}
  \Xmat_{ft,K} = \Amat_{f,K}\Smat_{ft,K}.
\end{equation}
We will refer to $\{\Xmat_{ft,K}\}_{f=1,t=1}^{F,T}$ as a \textit{multichannel, $K$-frame spectrogram}. Each time-frequency bin contains an $M\times K$ complex matrix.
The question then becomes: \textit{is there a non-constant function $g$ and a function $h$ such that $g(\Xmat_{ft,K})=h(\Amat_{f,K})$ for all $\Amat_{f,K}\in\mathbb{C}^{M\times K}$ and $\Smat_{ft,K}\in\mathbb{C}^{K\times K}$?} From now and until the end of this paper, we will assume that the number of sources is strictly lower than the number of microphones, \textit{i.e.} $K<M$. Under this assumption, an interesting candidate solution is $g=h=\operatorname{span}$, where $\operatorname{span}\hspace{-0.9mm}:\mathbb{C}^{M\times K}\rightarrow\operatorname{Gr}(K,\mathbb{C}^M)$ is the function associating a matrix to its column space. $\operatorname{Gr}(K,\mathbb{C}^M)$ is called a \textit{Grassmannian manifold}: elements of this set are $K$-dimensional linear subspaces of $\mathbb{C}^M$ \cite{griffiths1994principles,taseska2014subspace}. Assuming that the square matrix $\Smat_{ft,K}$ has linearly independent columns (this assumption is further discussed in Section \ref{subsec:conditions}), it acts as a change of basis from the column space of $\Amat_{f,K}$ to the column space of $\Xmat_{ft,K}$ in equation (\ref{eq:KTsource_model}). Therefore, $\operatorname{span}(\Xmat_{ft,K}) = \operatorname{span}(\Amat_{f,K})$ does not depend on $\Smat_{ft,K}$, and $\operatorname{span}$ possesses the desired properties to generalize RTFs.

However, the output values of $\operatorname{span}$ are not vectors but vector subspaces. These cannot be manipulated numerically. We thus need a way to map the Grassmannian manifold $\operatorname{Gr}(K,\mathbb{C}^M)$ to a numerical space. This is possible using a method known as \textit{Pl\"ucker embedding} \cite{griffiths1994principles}. The method was first introduced in the case $K=2$ and $M=4$ by Julius Pl\"ucker in 1865, and later generalized to any $K$ and $M$ values by Hermann Grassmann. Building on this, we propose a new transform for multichannel, multi-frame spectrograms. This transform applied to equation (\ref{eq:KTsource_model}) will yield an equation of the form (\ref{eq:1source_model}), allowing a generalization of RTFs to multiple sources. We shall name it the \textit{Pl\"ucker spectrogram transform} after the work of Julius Pl\"ucker.

\subsection{The Pl\"ucker spectrogram transform}
\label{subsec:plucker}
Let $\{\Xmat_{ft,K}\}_{f=1,t=1}^{F,T}$ be an $M$-channel $K$-frame spectrogram. We denote by $\Xmat_{ft,K|i_1,i_2,\dots,i_K}$ the $K\times K$ matrix formed by the $K$ rows of $\Xmat_{ft,K}$ with indexes $i_1,i_2,\dots,i_K$. Let $\xi(1),\dots,\xi(L)$ be the lexicographically-ordered list of cardinal-$K$ sublists of $\{1,\dots,M\}$ with $L=\binom{M}{K}$. We define the \textit{Pl\"ucker spectrogram transform of order $K$} as follows:
\begin{equation}
\label{eq:plucker}
 \mathfrak{p}_K(\Xmat_{ft,K})=\frac{1}{K!}\left(
  \begin{array}{c}
   \det(\Xmat_{ft,K|\xi(1)})\\
   \det(\Xmat_{ft,K|\xi(2)})\\
   \vdots\\
   \det(\Xmat_{ft,K|\xi(L)})\\
  \end{array}
  \right)\in\mathbb{C}^L.
\end{equation}
This transform applied to (\ref{eq:KTsource_model}) yields the following remarkable identity:
\begin{equation}
   \label{eq:identity}
   \mathfrak{p}_K(\Xmat_{ft,K}) = \mathfrak{p}_K(\Amat_{f,K})\det(\Smat_{ft,K}).
\end{equation}
This follows from the determinant property $\det(\Amat\Bmat)=\det(\Amat)\det(\Bmat)$ for square matrices $\Amat$ and $\Bmat$ of equal sizes. Interestingly, (\ref{eq:identity}) has the same form as equation (\ref{eq:1source_model}). In other words, the Pl\"ucker spectrogram transform changes an $M$-microphone observation of $K$ sources into an $\binom{M}{K}$-microphone observation of a single (compound) source. As a consequence, we have:
\begin{equation}
\label{eq:GRTF}
\rvect_{f,K} = \nu(\mathfrak{p}_K(\Xmat_{ft,K})) = \nu(\mathfrak{p}_K(\Amat_{f,K})).
\end{equation}
Therefore, $\rvect_{f,K}$ is a suitable generalization of RTFs to $K$ sources and $M$ microphones ($K<M$) using multiframe spectrograms. Namely, it verifies properties \textit{(I)}, (\textit{II}) and (\textit{III}), and for $K=1$, the RTF definition given in Section \ref{subsec:1source} is exactly recovered.

\subsection{Relation to subspace methods}
The proposed approach shares a lot of similarities with the so-called \textit{subspace methods} for sound source localization. A well-known example is the method MUSIC, which stands for MUltiple SIgnal Classification, \cite{schmidt1986multiple,argentieri2007broadband}. MUSIC starts by computing the covariance matrix of a multichannel signal in a given frequency band. An eigenvalue decomposition of this matrix is then performed, allowing to identify the \textit{signal subspace}, spanned by the principal eigenvectors, and the orthogonal \textit{noise subspace}, spanned by the remaining eigenvectors. As showed in Section \ref{subsec:multicase}, the signal subspace corresponds to the space spanned by the ATF, or equivalently the RTF vectors associated to the emitting sources, \textit{i.e.}, $\operatorname{span}(\Amat_{f,K})$. In contrast, RTF vectors are orthogonal to the noise subspace. Therefore, sound source directions are those whose associated RTF vectors have minimal projections onto the noise subspace. They are usually estimated by finding the smallest projections of a predefined set of RTF vectors.

Alternatively, in equation (\ref{eq:GRTF}), we introduce a new vector $\rvect_{f,K}$ which \textit{uniquely characterizes} the signal subspace $\operatorname{span}(\Amat_{f,K})$, using a minimal number of observations. This vector can thus be directly mapped to the spatial properties of all sources, provided that the associated mapping function is known. This mapping may either be directly obtained from a sound propagation model or learned from a predefined set of RTF vectors, as demonstrated in Section \ref{sec:results}. An intrinsic difference between this approach and MUSIC is that it does not require the estimation and decomposition of covariance matrices. On the other hand, it requires a mapping from generalized RTFs to multiple-source spatial characteristics, while MUSIC only requires single-source mappings.

\subsection{Conditions of applicability and properties}
\label{subsec:conditions}
Assuming that the normalizing function $\nu$ divides a vector by, \textit{e.g.}, its first entry, (\ref{eq:GRTF}) is only valid if $\det(\Xmat_{ft,K|\xi(1)})\ne0$. Using (\ref{eq:plucker}), (\ref{eq:identity}) and properties of the determinant, it follows that such singularity only occurs in the following situations:
\begin{itemize}
 \item If one or more sources are completely silent in all $K$ time frames $(t\dots t+K-1)$ at frequency $f$.
 \item If two or more sources are perfectly correlated over the segment, \textit{i.e.}, their absolute normalized cross-correlation is 1.
 \item If two or more sources have \textit{similar} spatial properties, \textit{i.e.} $\avect_{f,k}=\alpha\avect_{f,l}$ for some $\alpha\in\mathbb{C}, k\ne l$. This may occur if, \textit{e.g.}, they have identical directions in the free-field case.
 \item If the $K$ transfer functions and emitted signals are such that observations are linearly dependent, by coincidence.
\end{itemize}
Let us define audio sources as objects emitting distinguishable sounds from distinguishable locations. Then, the first three cases may be interpreted as a violation of the assumption that there are $K$ sources. The fourth case is harder to interpret, but it has a zero probability of occurrence assuming that distinguishable transfer functions and signals are mutually statistically independent. In other words, the proposed generalization of RTF is sound if the assumed number of sources $K$ is correct. If the actual number of sources $P$ at $(f,t)$ is less than $K$ then $\mathfrak{p}_K(\Xmat_{ft,K})=0$. If $P>K$, the desirable properties are no longer preserved.
A straightforward way to determine $P$ is to note that:
\begin{equation}
\label{eq:rank}
 P = \operatorname{rank}(\Xmat_{ft,K}) \textrm{ for } K>P.
\end{equation}
If $P<M$, $P$ can thus be deduced by successively calculating $\operatorname{rank}(\Xmat_{ft,K})$ for $K=1\dots M-1$.

\section{Simulated experiments}
\label{sec:results}
We test the potential of the proposed generalization of RTF for multiple sound-source localization (SSL). In what follows, spectrograms are computed on signals sampled at 8,000 kHz using 32 ms sliding windows with 50\% overlap. This results in $F=128$ positive frequencies and $T=64$ time frames per second of signal. We use a dataset of \textit{head-related transfer functions} (HRTFs) for the humanoid robot NAO. These HRTFs are simulated using a 3D model of the head in an anechoic environment and the boundary element method, as done in \cite{tourbabin2014theoretical}. Corresponding impulse responses have a maximal length of $10$ms. The subset $\mathcal{H}$ used contains $N=21$ HRTFs $\{\avect_f(\thetavect_n)\}_{f=1,n=1}^{F,N}\subset\mathbb{C}^M$ for the $M=4$ microphones placed on the head. Here $\Theta = \{\thetavect_1\dots\thetavect_N\}$ is a set of source directions with azimuth and elevations randomly picked in $[-180^\circ,180^\circ]$ and $[-10^\circ,10^\circ]$ respectively. From this dataset, the following \textit{generalized RTF} (GRTF) training sets are generated, for $K$=1 to 3:
\begin{align}         
  \mathcal{R}_K &= \bigl\{\nu(\mathfrak{p}_K([\avect_f(\thetavect_1), \dots, \avect_f(\thetavect_K)])); \nonumber \\
		&\hspace{7mm} \thetavect_1<\dots<\thetavect_K\in\Theta, f=1\dots F\bigr\} \nonumber   		
\end{align}
where the cardinality of $\mathcal{R}_K$ is $F\binom{N}{K}$. We then simulate all possible $M$-microphone mixtures of one to three white-noise sources coming from distinct directions in $\Theta$, by convolving random signals of one second duration with the HRTFs in $\mathcal{H}$. The minimum distance between distinct sources is $1^\circ$ in azimuth and $3^\circ$ in elevation. These mixtures are perturbed by additive Gaussian noise with 10 dB or 50 dB \textit{signal-to-noise ratios} (SNRs). The Pl\"ucker spectrogram transform of order $K$ (\ref{eq:plucker}) is then applied to all individual $K$-frame time segments of all these mixtures, where $K$ is the number of sources, assumed known. The $F$ GRTFs associated with the $F$ frequency bins at each segment are concatenated and compared to those of the corresponding training set $\mathcal{R}_K$, in terms of Euclidean distance. The set of $K$ directions minimizing this distance gives the estimated sound source directions. For $K=1$, $2$ and $3$, this respectively corresponds to approximately $1,300$, $26,000$ and $250,000$ localization tasks using time segments of length 32ms, 48ms and 64ms. The mean computational times per source per second of signal where respectively 81ms, 87ms and 436ms using our Matlab implementation on a conventional PC. Mean absolute azimuth localization errors obtained with this procedure are summarized in Table \ref{tab:results} (GRTF). 

\begin{table}
%\begin{wraptable}[6]{l}[.5\width+.5\columnsep]{6cm}
\centering
   \caption {\label{tab:results} {Mean absolute azimuth localization error using generalized RTFs on mixtures of 1 to 3 sources, with 10 or 50 dB signal-to-noise ratios.}}
   \begin{tabular}{|cc|ccc|}
      \hline
      \multicolumn{2}{|c|}{Number of sources}  & $1$ & $2$  & $3$  \\
      \hline
      \multicolumn{2}{|c|}{GRTF (SNR=50 dB)}  & $0.04^\circ$  & $0.68^\circ$  & $1.45^\circ$  \\                
      %\hline
      \multicolumn{2}{|c|}{GRTF (SNR=10 dB)} & $10.9^\circ$  & $17.5^\circ$  & $27.4^\circ$  \\                
      \hline     
   \end{tabular}
   \vspace{-3mm}
\end{table}

The results confirm that the proposed generalization of RTF captures spatial properties of sources under low noise level (50 dB SNR). However, performance is severely degraded for higher noise levels (10 dB SNR). 
While these results are only preliminary, they reveal two intrinsic benefits of the proposed approach. First, it can localize $K$ simultaneous sound sources using only $K$ spectrogram time frames. For $K=3$ and 50 dB SNR, $91\%$ of the 250,000 individual sources were perfectly localized using GRTFs on 64ms segments. This is impossible using methods such as MUSIC \cite{argentieri2007broadband}, where at least $M$ and typically more time frames are required to reliably estimate spatial covariance matrices. Second, the $K$ sound sources are jointly localized without using source separation, even though their spectra are strongly overlapping (white noise). This makes the method intrinsically efficient computationally, and contrasts with many existing multiple sound source localization methods, which rely on source separation \cite{lombard2009multidimensional,MandelWeissEllis10,deleforge2015acoustic}. These two features put forward GRTFs as a promising tool to efficiently localize multiple sound sources using short time windows. This ability may turn out to be critical, \textit{e.g.}, in realistic human-robot interaction scenarios where sound sources may be fast moving and computational resources are limited.

\section{Conclusion}
\label{sec:conclusions}
We proposed a natural way of generalizing relative transfer functions to $K$ sources using $K$ spectro-temporal observations, where $K$ is lower than the number of microphones. To the best of the authors' knowledge, this is the first study of this kind in signal processing. This work is mostly preliminary and theoretical. In the future, we plan an in-depth theoretical and empirical study of the noisy case, and an extension to natural sounds with sparse spectrograms such as speech. Moreover, several leads will be investigated to improve robustness to noise, \textit{e.g.}, estimating the number of sources, combining Pl\"ucker transforms of different orders and weighting time-frequency observations. Finally, the possibility of learning the mapping function from GRTFs to source directions will be investigated, following \cite{deleforge2015acoustic}. This would bypass the need for a comprehensive training set containing all possible combination of source positions.

\bibliographystyle{IEEEbib}
%\bibliography{refs}

\begin{thebibliography}{10}

\bibitem{parra2000convolutive}
Lucas Parra and Clay Spence,
\newblock ``Convolutive blind separation of non-stationary sources,''
\newblock {\em IEEE Transactions on Speech and Audio Processing}, vol. 8, no.
  3, pp. 320--327, 2000.

\bibitem{affes1997signal}
Sofiene Affes and Yves Grenier,
\newblock ``A signal subspace tracking algorithm for microphone array
  processing of speech,''
\newblock {\em IEEE Transactions on Speech and Audio Processing}, vol. 5, no.
  5, pp. 425--437, 1997.

\bibitem{dvorkind2005time}
Tsvi~G. Dvorkind and Sharon Gannot,
\newblock ``Time difference of arrival estimation of speech source in a noisy
  and reverberant environment,''
\newblock {\em Signal Processing}, vol. 85, no. 1, pp. 177--204, 2005.

\bibitem{laufer2013relative}
Bracha Laufer, Ronen Talmon, and Sharon Gannot,
\newblock ``Relative transfer function modeling for supervised source
  localization,''
\newblock in {\em IEEE Workshop on Applications of Signal Processing to Audio
  and Acoustics (WASPAA), 2013}. IEEE, 2013.

\bibitem{deleforge2015acoustic}
Antoine Deleforge, Florence Forbes, and Radu Horaud,
\newblock ``Acoustic space learning for sound-source separation and
  localization on binaural manifolds,''
\newblock {\em International journal of neural systems}, vol. 25, no. 1, pp.
  1--21, 2015.

\bibitem{benesty2001advances}
Jacob Benesty, Tomas G{\"a}nsler, Dennis~R Morgan, M~Mohan Sondhi, Steven~L
  Gay, et~al.,
\newblock {\em Advances in network and acoustic echo cancellation},
\newblock Springer, 2001.

\bibitem{gannot2001signal}
Sharon Gannot, David Burshtein, and Ehud Weinstein,
\newblock ``Signal enhancement using beamforming and nonstationarity with
  applications to speech,''
\newblock {\em IEEE Transactions on Signal Processing}, vol. 49, no. 8, pp.
  1614--1626, 2001.

\bibitem{markovich2009multichannel}
Shmulik Markovich, Sharon Gannot, and Israel Cohen,
\newblock ``Multichannel eigenspace beamforming in a reverberant noisy
  environment with multiple interfering speech signals,''
\newblock {\em IEEE Transactions on Audio, Speech, and Language Processing},
  vol. 17, no. 6, pp. 1071--1086, 2009.

\bibitem{reindl2013geometrically}
Klaus Reindl, S~Markovich-Golan, Hendrik Barfuss, Sharon Gannot, and Walter
  Kellermann,
\newblock ``Geometrically constrained trinicon-based relative transfer function
  estimation in underdetermined scenarios,''
\newblock in {\em IEEE Workshop on Applications of Signal Processing to Audio
  and Acoustics (WASPAA), 2013}. IEEE, 2013.

\bibitem{duda1997elevation}
Richard~O Duda,
\newblock ``Elevation dependence of the interaural transfer function,''
\newblock {\em Binaural and spatial hearing in real and virtual environments},
  pp. 49--75, 1997.

\bibitem{blauert2013technology}
Jens Blauert,
\newblock {\em The technology of binaural listening},
\newblock Springer, 2013.

\bibitem{griffiths1994principles}
Phillip Griffiths and Joseph Harris,
\newblock {\em Principles of algebraic geometry},
\newblock John Wiley \& Sons, 1994.

\bibitem{taseska2014subspace}
Maja Taseska and Emanuel~AP Habets,
\newblock ``A subspace-based perspective on spatial filtering performance with
  distributed and co-located microphone arrays,''
\newblock in {\em ITG Fachtagung Sprachkommunikation}. VDE, 2014.

\bibitem{schmidt1986multiple}
Ralph~O Schmidt,
\newblock ``Multiple emitter location and signal parameter estimation,''
\newblock {\em Antennas and Propagation, IEEE Transactions on}, vol. 34, no. 3,
  pp. 276--280, 1986.

\bibitem{argentieri2007broadband}
Sylvain Argentieri and Patrick Danes,
\newblock ``Broadband variations of the {MUSIC} high-resolution method for
  sound source localization in robotics,''
\newblock in {\em IEEE/RSJ International Conference on Intelligent Robots and
  Systems (IROS), 2007}. IEEE, 2007, pp. 2009--2014.

\bibitem{tourbabin2014theoretical}
Vladimir Tourbabin and Boaz Rafaely,
\newblock ``Theoretical framework for the optimization of microphone array
  configuration for humanoid robot audition,''
\newblock {\em IEEE Transactions on Audio, Speech and Language Processing},
  vol. 22, no. 12, pp. 1803--1814, 2014.

\bibitem{lombard2009multidimensional}
Anthony Lombard, Tobias Rosenkranz, Herbert Buchner, and Walter Kellermann,
\newblock ``Multidimensional localization of multiple sound sources using
  averaged directivity patterns of blind source separation systems,''
\newblock in {\em IEEE International Conference on Acoustics, Speech and Signal
  Processing (ICASSP), 2009.} IEEE, 2009, pp. 233--236.

\bibitem{MandelWeissEllis10}
Michael~I. Mandel, Ron~J. Weiss, and Daniel P.~W. Ellis,
\newblock ``Model-based expectation-maximization source separation and
  localization,''
\newblock {\em IEEE Transactions on Audio, Speech and Language Processing},
  vol. 18, no. 2, pp. 382--394, 2010.

\end{thebibliography}

\end{document}